\documentclass[preprint]{aastex6}

\usepackage{amssymb}
\usepackage{soul}
\usepackage{graphicx}
\usepackage{amsmath}
\usepackage{url}

\begin{document}
\sloppy

\title{TESSVisibility -- When was my favorite star or asteroid observed by TESS?}
\shorttitle{TESSVisibility}

\author{Andr\'as P\'al\altaffilmark{1,2,3}}

\altaffiltext{1}{Konkoly Observatory, Research Centre for Astronomy and Earth Sciences, H-1121 Budapest, Konkoly Thege Mikl\'os \'ut 15-17, Hungary}
\altaffiltext{2}{E\"otv\"os Lor\'and University, Institute of Physics, P\'azm\'any P\'eter s\'et\'any 1/A, 1117 Budapest, Hungary}
\altaffiltext{3}{MIT Kavli Institute for Astrophysics and Space Research, 70 Vassar Street, Cambridge, MA 02109, USA}

\email{apal@konkoly.hu}

\begin{abstract}
While Transiting Exoplanet Survey Satellite (TESS) covers a considerable area of the sky during 
routine observations and the pointing schedule is easy to follow, it is not obvious to retrieve the
current and/or predicted visibility of a bulk amount of objects, considering both stationary and moving 
Solar System targets like asteroids or comets. 
The program \texttt{tessvisibility} is a small piece of highly portable 
code implemented in both C an UNIX shell, providing functionalities for 
such bulk retrievals at the accuracy of a TESS pixel. This accuracy 
includes the gaps between the focal plane CCDs, the gaps between the cameras 
as well as at the sector-level treatment to obtain visibility information. 
\end{abstract}


\section{Background}
\label{sec:background}

Launched in April 2018, {\it Transiting Exoplanet Survey Satellite} 
\citep[{\it TESS},][]{ricker2015} performs routine operations
since 25 July, 2018 by observing $24^\circ\times 96^\circ$ sections of 
the sky called {\it sectors} in anti-solar directions. 
This large field-of-view (FoV) is provided by four identical cameras, 
each having a fast lens and an
array of four frame-transfer CCDs in the focal plane. Therefore, the 
instrument FoV is effectively assembled from $16$ continuous parts, each
having a size of approx. $12^\circ\times12^\circ$. Because of this unique combination
of the fast optics and the focal plane arrangement, the optical axes even do 
not fall on silicon and significant amount of optical distortions are also 
needed to be taken into account when one computes the visibility of objects. 
Even though, the gaps between the CCDs are equivalent to $60-100$ pixels,
therefore, a na\"{\i}ve check would falsely confirm the visibility of 
an object with a chance of $8-10\%$ just by checking the full per-camera
FoV of $24^\circ\times24^\circ$. 

\section{Implementation and availability}
\label{sec:implementation}

The core functionality of the \texttt{tessvisibility} toolchain is to 
perform efficiently the astrometric conversion between J2000 celestial 
coordinates and CCD pixel coordinates in the frame used by the publicly 
available raw and calibrated full-frame images (FFIs). Fundamentals and
the sub-steps of the process follow the \cite{pal2020} pipeline implemented
for extracting asteroid photometry on TESS FFIs. Namely, a) the spacecraft
boresight pointing information is converted into per-camera pointing by simply
taking a fixed $24^\circ$ separation of the intended optical axes, b) 
a gnomonic projection is applied in the focal plane, followed by a three-term
radial Brown-Conrady distortion model, c) a third-order per-CCD polynomial 
is applied to derive the final pixel coordinates. The employed CCD geometry 
is in accordance with Fig.~4.3 of \cite{vanderspek2018}, i.e. a visible
object should appear between $(44,0)\le (x,y) \le (2092,2048)$ -- in 
other words, the underclock, overclock and calibration areas are excluded. 
While the geometric parameters of the final astrometric solutions are 
kept fixed in the steps a) and b), the polynomial
coefficients were derived empirically using an astrometric fit to 
the GAIA DR2 catalogue \citep{gaia2016,gaia2018} by involving the tasks
of the FITSH package \citep{pal2012}. 
Based on the analysis of the individual per-sector astrometric solutions, 
the hard-wired polynomial coefficients provide a sub-pixel accuracy for the 
overall computation of the apparent positions. 

Besides the simple conversion between J2000 and plate coordinates,
the C version of \texttt{tessvisibility} is capable to perform 
bulk conversation and visibility check on a user-defined input list 
of coordinates as well as time series
of astrometric positions of moving objects by reading simple textual files. 
In this manner, \texttt{tessvisibility} follows the philosophy of the classic 
UNIX pipelines and can be used in conjunction with various 
command line utilities and any kind of programs implementing such features.
Both the C and UNIX shell (\texttt{bash}) version of the code has 
no dependencies, all of the functionalities are implemented within the
source code. The only exception is the sector boundary retrieval for the
\texttt{bash} version, which relies on the FITSH 
package\footnote{\url{https://fitsh.net/}} \citep{pal2012} pre-installed. 
Both implementations have a built-in support for the publicly available 
pointing data\footnote{\url{https://tess.mit.edu/observations/}} up to
the end of the 4th year of the mission (i.e. up to 2022-08-31). In addition,
the C version also features a {\it custom pointing} mode where the 
boresight pointing is parameterized via an additional argument for the 
program. In the analysis of ephemerides (i.e. in the ``moving object''
mode, see above), the user should also supply a corresponding timespan 
for this particular custom pointing as a mandatory command line argument. 

The aforementioned implementations of the \texttt{tessvisibility} code
are available from the Konkoly Observatory 
archives\footnote{\url{https://archive.konkoly.hu/pub/util/tess/tessvisibility/}}.
Both implementations are capable to resolve object names using the CDS/Sesame
service\footnote{\url{http://cds.u-strasbg.fr/cgi-bin/Sesame}}. 

\facilities{TESS \citep{ricker2015}, \textit{Gaia} DR2 \citep{gaia2018}} 
\software{FITSH \citep{pal2012}, EPHEMD \citep{pal2020}}

\vspace*{2mm}

\begin{acknowledgements}
The author would like to thank Attila B\'odi for checking the MacOS portability
of the source code. This work was partially supported by the GINOP-2.3.2-15-2016-00033 
project which is funded by the Hungarian National Research, 
Development and Innovation Fund together with the  European Union. Additional
support was received from the Hungarian National Research, Development and Innovation Office
(NKFIH) grant and K-125015.
\end{acknowledgements}


{}

\end{document}